# "Ground Truth" calibration for the JEM-EUSO Mission

J.H. Adams, Jr.[a], M.J. Christl[b], S.E. Csorna[c], F. Sarazin[d] and L.R. Wiencke[d] for the JEM-EUSO Collaboration

[a] University of Alabama in Huntsville, Huntsville, AL 35805, USA, jha0003@uah.edu
[b] NASA Marshall Space Flight Center, Huntsville, AL 35812, USA, mark.christl@nasa.gov
[c] Vanderbilt University, Nashville, TN 37235, USA, steven.e.csorna@vanderbilt.edu
[d] Colorado School of Mines, Golden, CO 80401 USA, lwiencke@mines.edu

**Abstract**

The Extreme Universe Space Observatory is an experiment to investigate the highest energy cosmic rays by recording the extensive air showers they create in the Earth's atmosphere. This will be done by recording video clips of the development of these showers using a large high-speed video camera to be located on the Japanese Experiment Module of the International Space Station. The video clips will be used to determine the energies and arrival directions of these cosmic rays. The accuracy of these measurements depends on measuring the intrinsic luminosity and the direction of each shower accurately. This paper describes how the accuracy of these measurements will be tested and improved during the mission using a global light system consisting of calibrated flash lamps and lasers located deep in the Earth's atmosphere.

## 1. Introduction

The Extreme Universe Space Observatory on the Japanese Experiment Module (JEM-EUSO) of the International Space Station (ISS) is shown in Fig. 1. It will investigate the most powerful cosmic accelerators in the Universe which are the sources of extreme energy cosmic rays (EECRs), those with energies $>10^{20}$ eV (Medina-Tanco et al., 2012).

JEM-EUSO is a large wide-angle and high-speed video camera designed to capture video clips of the extensive air showers (EASs) created by EECRs striking the atmosphere (Ricci et al., 2012). These video clips will be used to reconstruct the energies and arrival directions (Bertaina et al., 2011 and Casolino, M. et al., 2011) of the EASs. To reconstruct the energy, the intrinsic luminosity of the EAS must be measured (Mernik et al., 2011) and to reconstruct the arrival direction of the EECR, the direction of the EAS in celestial coordinates must be accurately reconstructed. To reconstruct the flux of EECRs, the trigger efficiency must be known. These are the three key observables for identifying the sources of the highest particles in the universe and for evaluating the scientific performance of the pioneering JEM-EUSO instrument.

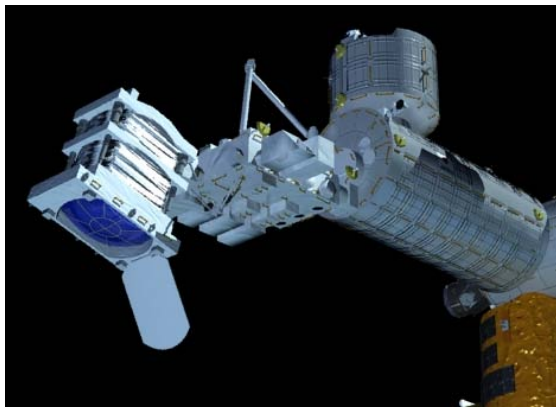

Figure 1: An artist's concept of the JEM-EUSO camera installed on the International Space Station (courtesy of RIKEN)

The sources of error for JEM-EUSO include:

a) uncertainties in the atmospheric conditions,
b) background light levels,
c) timing uncertainties,
d) tilt angle of JEM-EUSO,
e) pointing errors (due to the attitude instability of the ISS),
f) temperature variations and
g) the age of the instrument.

Making these measurements accurately requires careful calibration and monitoring of the instrument and characterization of atmospheric conditions as a function of time throughout the mission. An onboard monitoring system of UV-LEDS is operated during daylight segments of the flight when the aperture door is closed. This subsystem monitors the relative variation of individual pixels of the focal surface and the transmission in the lenses during the mission (Sakaki, Naoto et al., 2011). The atmosphere conditions in the vicinity of an EAS are characterized by an onboard Atmospheric Monitor (AM) employing a LIDAR and an IR camera (Takayuki et al., 2012 and Shibata et al., 2011) and by a special slow acquisition mode used by JEM-EUSO to characterize the background photon level, altitude and opacity of any low-lying clouds present (Garino et al., 2011). This data, together with a model of the atmosphere and the location and inclination of the EAS, are used to correct for absorption and scattering losses suffered by the fluorescence and Cherenkov photons in each EAS (Sáez Cano et al., 2012). In addition, the attitude of JEM-EUSO is monitored continuously during the mission. These measurements and calibrations are used to reconstruct the intrinsic luminosity of the EASs and their direction. They also are used to adjust the trigger conditions in order to maintain the trigger efficiency.

To check the accuracy with which the intrinsic luminosity and the EECR arrival direction are determined and to measure the trigger efficiency, reference sources in the atmosphere must be used. The proposed Global Light System (GLS) will provide these reference sources to check these three key parameters.

## 2. The Global Light System (GLS)

The GLS is a worldwide network, operated remotely, that combines ground-based xenon flash lamps and lasers to provide a practical and cost-effective means of validating the three key parameters. The GLS will generate benchmark optical signatures in the atmosphere with characteristics similar to the optical signals of EASs. But unlike air showers, the number, rate, intrinsic luminosity, precise time and direction (in the case of the lasers) are known. The calibrated signals from the GLS units and the measurements of these signals made by JEM-EUSO will be compared. Throughout its mission, JEM-EUSO will reconstruct the positions and pointing directions of the lasers and the intrinsic luminosities of the lasers and flash lamps. The laser shots, which mimic EAS tracks with known luminosity profiles, will be used to monitor JEM-EUSO's trigger efficiency. The lasers will produce tracks that point in the directions of known astrophysical objects. The track directions as reconstructed from the video clips of these laser shots will be compared to the known directions of the laser shots. From these comparisons, the accuracy of the measured EECR arrival directions can be estimated. The xenon flash lamps provide UV light flashes of known intrinsic luminosity. These flashes will be recorded by JEM-EUSO and their intrinsic luminosities will be reconstructed. By comparing the reconstructed luminosities with the calibrated luminosities of these flashers, the accuracy of the energy measurements of EECRs can be determined.

The GLS provides the means to adjust the JEM-EUSO instrument so that the key parameters can be accurately determined. Using the comparisons between the ground truth provided by the GLS and the reconstructed parameters from the on-orbit observations, the reconstruction algorithms and the trigger can be adjusted. In addition, the GLS provides point sources of light that can be used to check and adjust the focus of the JEM-EUSO optics.

## 2.1 The GLS Installations

Table 1: The Global Light System will include three types of units.

| GLS unit | Sources | # |
|---|---|---|
| GLS-X | Xenon Flash-lamps (XF) | 6 |
| GLS-XL | XF and Laser | 6 |
| GLS-AXHL | XF and Horizontal Laser(airborne) | 1 |

The GLS units are described in Table 1. There will be twelve ground-based units strategically placed at sites around the world. Six will have both xenon flashers and a steerable laser (GLS-XL). The remaining six will have only xenon flashers. There will also be one airborne GLS-XL unit operated out of NASA's Wallops Flight Facility. The ground-based sites will be chosen for their low background light. These sites will also be at altitudes that are above most of the planetary boundary layer and typical of the deeper EECR shower maxima. The placement of GLS units will cover a broad range of terrestrial conditions (land, open sea), geographical locations, latitudes, and altitudes. A list of candidate sites is provided in the Table 2.

In one orbit, JEM-EUSO views an area of $1.6 \times 10^7$ km$^2$. Each day JEM-EUSO sweeps over an area of $2.5 \times 10^8$ km$^2$ between $\pm 51.6°$ latitude. On average, a GLS site will be over-flown in darkness and with favorable atmospheric conditions (based on 10% duty cycle for a ground based station, see Shinozaki, et al., 2011) each day of the mission. This will provide a data base that will be sensitive to seasonal variations at the locations. During an over-flight, the GLS unit will produce calibrated signals repeatedly while in JEM-EUSO's Field of View (FoV). A GLS unit will be within the FoV for 56 seconds on average. In addition, an airborne GLS unit will be mounted inside a P3B Orion aircraft and deployed monthly from Wallops Flight Facility/NASA (WFF) over the open ocean for under flights of JEM-EUSO at selected altitudes. Spare GLS units will be ready for deployment as needed to maintain operations

Table 2: Candidate sites for the land-based GLS units.

| Location | Latitude | Elevation |
|---|---|---|
| Jungfraujoch (Switzerland) | 47°N | 3.9 km |
| Mt. Washington (NH, USA) | 44°N | 1.9 km |
| Alma-Ata (Kazakhstan) | 44°N | 3.0 km |
| Climax (CO, USA) | 39°N | 3.5 km |
| Frisco Peak (UT, USA) | 39°N | 2.9 km |
| Mt Norikura (Japan) | 30°N | 4.3 km |
| Mauna Kea (HI, USA) | 20°N | >3.0 km |
| Nevado de Toluca (Mexico) | 19°N | 3.4 km |
| Chacaltaya (Bolivia) | 16°S | 5.3 km |
| La Reunion (Madagascar) | 21°S | 1.0 km |
| Cerro Tololo (Chile) | 30°S | 2.2 km |
| Sutherland (South Africa) | 32°S | 1.8 km |
| Pampa Amarilla (Argentina) | 35°S | 1.4 km |
| South Island (New Zealand) | 43°S | 1.0 km |

throughout the mission

The GLS is developed using Commercial Off-The-Shelf (COTS) components, assembled and tested at the University of Alabama in Huntsville, Colorado School of Mines and the Marshall Space Flight Center. First a prototype GLS-XL unit will be designed, fabricated and tested. This unit will be used to support a high-altitude balloon flight of a proto-type JEM-EUSO telescope, planned for 2014. The experience gained during the sub-orbital flight will be used to improve the GLS design so it can support the JEM-EUSO spaceflight mission.

**2.2 The Xe-flasher (GLS-X)**

Reconstruction of intrinsic luminosity of the EASs from JEM-EUSO measurements can be monitored directly with flash lamps on a daily basis (Adams et al., 2003; Adams et al., 2012). All GLS stations will include four flash lamps, filtered to closely match the three primary UV fluorescence lines (Arqueros et al, 2008a and 2008b; and Ave et al., 2008 and 2007) at 391 nm, 357 nm, and 337 nm (see Fig. 2) and a Schott BG3 wide band-pass filter, identical to that covering the focal plane detector in JEM-EUSO.

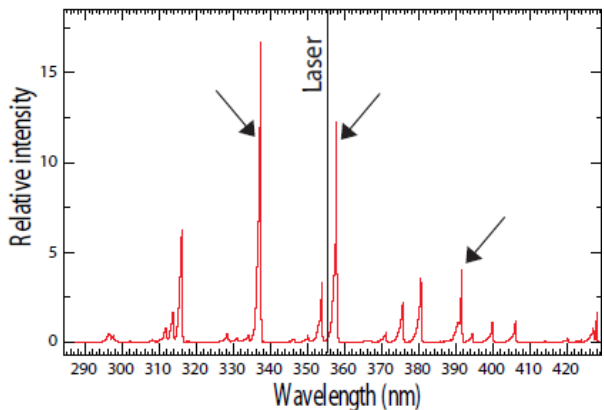

Figure 2: The measured fluorescence spectrum (Ave et al., 2007): The wavelength of the GLS flashers will be filtered to produce light at the lines indicated (arrows). The wavelength of the GLS lasers is also shown by the vertical black line. This figure is adapted from Ave et al. (2007).

The critical functional requirement is a high reliability light flash. Our GLS concept incorporates four individual UV-flashers (Hamamatsu L6604), three with line-filters and one with the broad band BG3 filter, that act as a set of "standard candles". The flashers are fitted with a mechanical aperture and clear glass outer window, followed by a glass diffuser and one of the optical filters (see Fig. 3). Each flasher includes other ancillary components to meet the performance specifications. These include a trigger module, a cooling jacket, capacitors for the initial and primary flashes and a power supply. The L6604 has a highly stable output of <3% (standard deviation) from flash-to-flash and a long stable lifetime with more than $10^7$ flashes and less than 3% degradation over the lifetime of the mission (ref: Hamamatsu Data Sheet). The light pattern from each flash is smoothly distributed over a wide field as shown in Fig. 4. The flash intensity is nearly uniform over a field of view that is >60°. The entrance aperture of JEM-EUSO limits the viewing angle at 30° from the nadir, well within the uniform angular range of the flash lamps. These key performance parameters have been verified in laboratory testing.

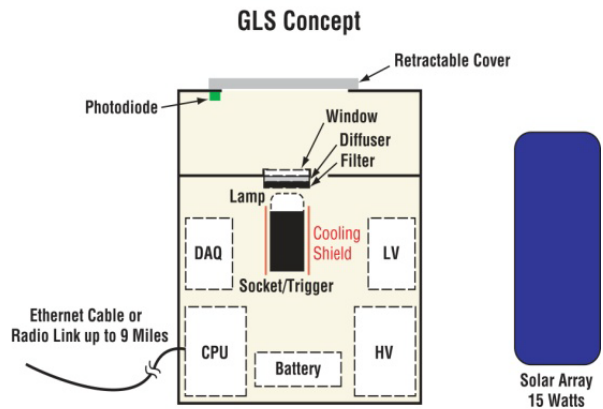

Figure 3: The GLS-X configuration showing the components of the system.

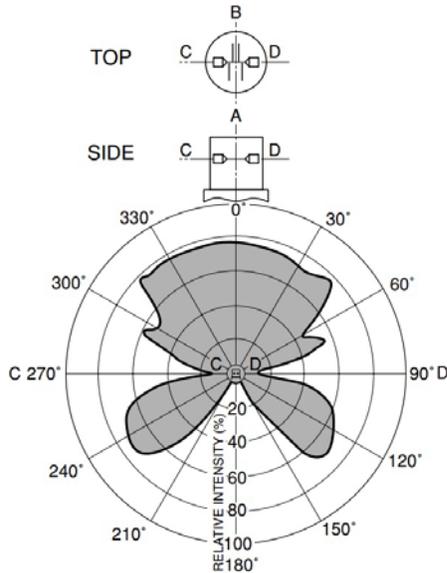

Figure 4: An example the Xe flash lamp intensity distribution (taken from Hamamatsu, 2005).

To better match the timing characteristics of a vertical EAS as seen by JEM-EUSO, the light pulse are stretched to >10 microseconds. This exceeds the minimum acceptable duration for the on-board trigger algorithm and enables JEM-EUSO to detect the GLS flashes without requiring a dedicated trigger. The capacitor and voltage are the principal parameters that control the timing and intensity characteristics of the flashes. We have investigated the performance of the flash lamps for various capacitor values and have succeeded in stretching the pulse while maintaining the stability and repeatability of the light flashes. We verified the performance in the laboratory by analyzing 100 flashes as shown in Fig 5. In this figure the intensities of the flashes have been converted to photo-electrons and scaled to the levels that would be detected by the JEM-EUSO instrument. This is done by estimating the losses for a clear nighttime atmosphere and the instrument efficiency factors. Each flash lasts ~24 μs. As shown in Fig. 5, the intensities of the flashes were measured by integrating the light over successive 2.33 microsecond intervals (similar to the duration of the Gate Timing Units (GTUs) planned for JEM-EUSO) spanning the flash duration. The standard deviations of the measured intensities for the 100 pulses in nine successive GTUs are shown by the error bars in Fig. 5. These deviations are significantly smaller than the variation expected due to photoelectron statistics for the signal recorded at JEM-EUSO; typically 30% smaller. The data in each pixel for successive GTUs will be summed to determine the total signal detected from each GLS flash.

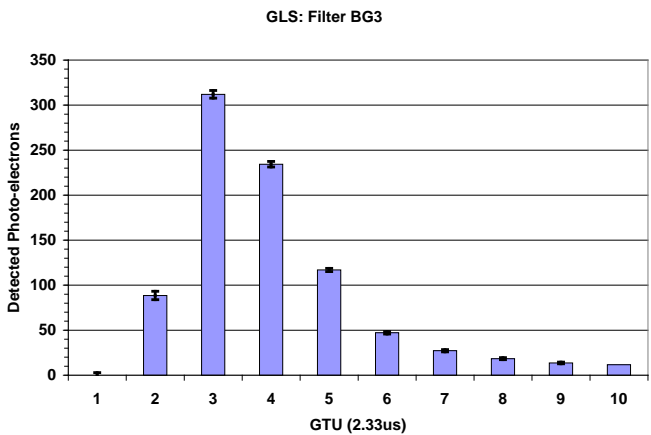

Figure 5: Measurements of the intensities of 100 flashes of a Xe flash lamp. Each flash lasts ~23 μsec. The measurement of each flash was made by integrating the intensity of the flash over 10 successive time intervals of 2.33 μsec each covering the duration of the flash. The vertical bars show the average intensity of the 100 flashes in each time interval. The error bars show the repeatability in each time interval. The intensity on the abscissa is presented in units of photoelectrons and scaled to the estimated intensities of the flashes as observed at JEM-EUSO under clear sky conditions.

The GLS flashers are housed in a weather tight enclosure that includes the flasher and its ancillary components, a photo-diode and data acquisition system, a mechanical shutter and motor, a Single Board Computer (SBC), solar array and battery (if needed) and a Remote Communication Element (RCE), see Fig. 3. The RCE provides communication capability with the Central Operations Center (COC) located at Marshall Space Flight Center to upload and download parameters for the GLS and to retrieve the data needed to determine its operational status. The internet connection for the RCE will depend on the on-site resources available. Three options will be available: a satellite link (through Iridium or OrbComm), a wireless router or a land-line. The SBC controls the operations of the GLSs. It performs housekeeping functions and reports on the status of the GLS to the COC. The GLS flash units will be mounted on top of steel poles to minimize potential interferences from the surrounding local environment. Discussions with personnel in the US forestry service experienced with remote sensing operations have been used to develop our initial plans for the sites. Members of the international JEM-EUSO Collaboration will provide assistance with identifying candidate sites for the GLS's and assistance in identifying and communicating with local authorities to assure the operation is free from any interference prior to selecting a site. Table 2 lists several candidate sites that meet the key requirements for the GLS: high elevation and low light pollution.

## 2.3 The GLS-XL

The laser subsystem of the GLS-XL leverages the successful experiences of the pioneering Fly's eye (Baltrusaitis et al., 1985 and Bird et al., 1995), HiRes (Wiencke et al., 1999 and Cannon et al., 2003) and Auger (Wiencke et al., 2011). The benchmark data from the automated remote lasers (Fick et al., 2006) in the Auger Fluorescence Telescope (FD) (see Abraham et al., 2010a) FoV have proven critical both for measuring the performance of the FD (Abraham et al., 2010b) and for realizing the science program (Abraham et al., 2010c). A 5 mJ 355nm laser track from a pulse shot across the field of view, and a $10^{20}$ eV air shower track look similar when viewed by the Auger florescence detector (Wiencke et al., 2009), see Fig. 6. Air shower tracks and laser shots will also look similar when viewed from JEM-EUSO. In the case of JEM-EUSO, the portions of both tracks that extend into the upper part of the troposphere will be elongated because the air is less dense.

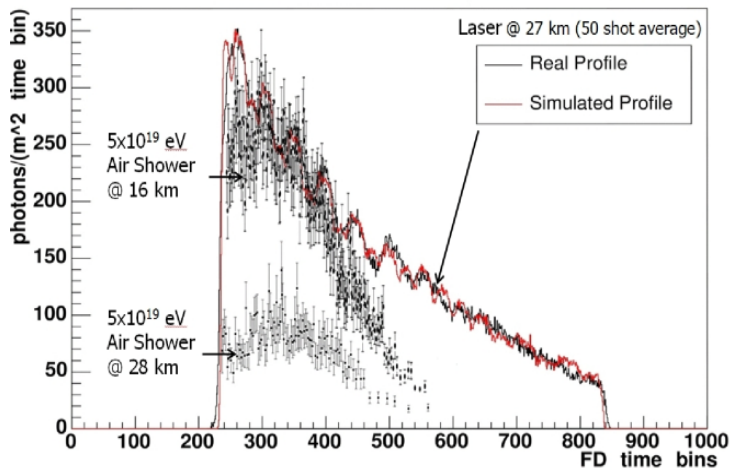

Figure 6: Longitudinal profiles of vertical laser shots and near-vertical cosmic ray air showers recorded by the Pierre Auger fluorescence detector from a horizontal distance of 27 km. One time bin is 100 ns. The viewed laser track is 12 km long in this example. The downward going EAS profiles have been "flipped" so that the left edge of all profiles match at the bottom of the FD field of view. This figure is taken from Wiencke (2009).

By adjusting the energy of the laser, the onboard JEM-EUSO trigger threshold for track-like optical signatures will also be tested. This technique is used at the Auger experiment (Fig. 6) from a GLS-X based on laboratory measurements. Lasers can also be aimed at EECR potential sources over the full sky. A selection of 20 sources could include, for example, the galactic center, Cen-A, Virgo, and other objects of astrophysical interest. A sky map of reconstructed laser track directions will be accumulated over the mission. The clusters of points and their spread about the directions of the targeted sources will provide a simple but comprehensive validation of the absolute EAS pointing accuracy reconstruction by JEM-EUSO. Included in this test is the correct transfer of time stamps from a precise clock on the ISS through the JEM-EUSO hardware and the onboard and ground based data analysis chains. During clear periods when 355 nm light propagation in the atmosphere between the GLS site and the ISS can be described to the few percent level using molecular scattering alone (i.e. aerosol optical depth << molecular optical depth), it will be possible to test intrinsic luminosity with both the direct Xe flasher light (point source) and the scattered UV laser light (track source). These clear periods can be identified by comparing the pattern of Xe flasher measurements made at different viewing angles as JEM-EUSO passes over to the predicted pattern based on molecular scattering and measurements made with the JEM-EUSO IR camera and onboard laser.

Table 3: GLS-XL laser parameters.

| Parameter | Spec |
| --- | --- |
| Wavelength | 355 nm |
| Energy/pulse | 1-10 mJ |
| Pulse Width | 10 ns |
| Energy Calib. (relative) | 3% |
| Energy Calib. (absolute) | <10% |
| Pointing (relative) | 0.02 deg. |
| Pointing (absolute) | 0.25 deg |
| Timing (Absolute) | 50 ns |

The GLS-XL is based on the design of a similar system in use at Auger for over 15 years. The specifications of the GLS-XL are listed in Table 3. The laser will use a frequency tripled YAG. Proven models are in use at HiRes (Wiencke et al., 1999 and Cannon et al., 2003) and Auger. The Auger system includes a BigSky (Quantel) compact folder resonator and a Centurion. The optics will include harmonic separator mirrors to remove residual primary and secondary harmonics. The net polarization of the beam will be randomized so that the atmosphere scatters the same amount of light azimuthally about the beam direction. The steering mechanism will use COTS controller and two orthogonal rotational stages following a design used by Auger and Hires. The steering mirrors will be coated for 355 nm and 632nm so that they can be aligned using an inexpensive leveling tool which employs a He-Ne laser. Since the laser beam will be far from JEM-EUSO, its beam will be much narrower than JEM-EUSO's resolution near in the lower atmosphere. Consequently, the beam diameter can be expanded significantly to avoid interference with aviation and other environmental impacts. The relative energy of each laser pulse will be measured by directing a small fixed fraction of the beam into a pyroelectric energy probe. When the steering mechanism is in the parked, (downward) position, the full beam energy will be measured by a second pyroelectric energy probe mounted inside the shelter to obtain an absolute energy calibration. To facilitate the identification of laser data in the JEM-EUSO data stream, the laser will be triggered at precise times relative to the GPS second and the times of each shot recorded locally. The system will include wind, rain, and temperature sensors and a webcam. The entire system will be computer-controlled using a SBC. The system will be connected to the internet through a firewall, possibly via wireless link. The system will be operated and reprogrammed remotely.

The system will be housed in an industrial enclosure/cabinet that is commercially available, temperature-controlled and designed for outdoor use. The steering head will be mounted on top of the enclosure and the laser, optics and controls will be mounted inside. Past experience has shown that these laser systems are the most stable, and thus provide the most useful benchmark data, when their temperature is controlled (20+/- 5° C) and the optics are kept free of dust. The optical enclosure and beam path through the steering mirrors will be sealed to minimize dust contamination. This compact arrangement, making use of a COTS enclosure is preferred after long experience using custom modified "connex" shipping containers and other structures. Using a compact COTS outdoor enclosure represents a departure from past systems that were housed in much larger modified shipping containers or custom structures. Consequently, it will be necessary to design and build a prototype and test an entire unit extensively, including the installation and alignment procedures, before finalizing the design and beginning fabrication of the 6 GLS-XL system planned for the JEM-EUSO mission.

## 3. Mission Operations

The GLS-to-EUSO interface consists of a trigger pattern that includes the range of intensities, wavelengths and timing characteristics for the GLS.

During an over-flight, a GLS-X will emit a fixed cadence of flashes with a range of wavelengths and intensities. The laser shots from a GLS-XL unit will also have a fixed cadence but a range of intensities and directions. During the mission, observations will be planed routinely. These plans will be uploaded every few days to the ISS, transferred to EUSO and executed autonomously. The GLS flashes and tracks will be processed by the same onboard processing as EAS triggered events and included in the telemetry stream downloaded to the JAXA operation center and transferred to the JEM-EUSO science center. This data also includes data generated by the AM system and the instrument housekeeping data.

The ISS ground track speed is 7.7 km/s. An individual pixel on the focal surface in JEM-EUSO has a footprint on the Earth's surface of 460 x 460 $m^2$ near the nadir for an ISS altitude of 350 km. As a result, an individual GLS site will be in the JEM-EUSO FoV on average for 56 seconds. The average pixel crossing time is 50 milliseconds so, on average, 1100 pixels will view a GLS site during a single over-flight by JEM-EUSO. The GLS-X operates at 10 Hz. A GLS-XL will alternate laser and flasher pulses for a total rate of 20 Hz providing a large number of measurements during a single JEM-EUSO over-flight. For each trigger, the AM system is automatically activated and acquires an IR camera image and a LIDAR shot at the location of the GLS (Neronov et al., 2011).

The xenon flasher is capable of delivering a signal to JEM-EUSO, under clear sky conditions of >800 photoelectrons per flash (Adams et al., 2003). The laser is capable of delivering a signal at JEM-EUSO of up to ~500 to 1000 photoelectrons per track under similar conditions. These signal levels can be adjusted to levels that match the JEM-EUSO trigger and photon counting systems. The GLS will emit multiple flashes during each over-flight. The measurements of these flashes onboard JEM-EUSO and the calibration data from the GLS xenon flasher will be analyzed to determine the UV attenuation of the atmosphere with an accuracy of a few percent. The UV attenuation measurement will be compared with the atmospheric attenuation estimated by the AM system onboard JEM-EUSO. From this comparison we will learn how accurately the AM system estimates atmospheric attenuation. The AM system attenuation estimates are used to reconstruct the intrinsic luminosity of EASs and hence the energy of the EECR so the measurements during over flights of GLS units allows us to learn how accurately JEM-EUSO measures the energies of EASs.

The deployment of the GLS will occur prior to the launch of JEM-EUSO. Twelve ground based sites will be selected and the GLS units will be installed and tested. Each site includes a microprocessor or SBC with internet access, enabling remote management. Operations will be conducted remotely from the COC. University of Alabama in Huntsville will be responsible for programming the xenon flashers and monitoring their performance. The Colorado School of Mines will be responsible for programming the lasers and monitoring their performance. The GLS flash intensities will be pre-loaded for each scheduled over-flight, based on predicted atmospheric conditions.  The pattern of laser shots will be programmed in advance and will include shots fired at potential EECR sources. After each over-flight, GLS housekeeping data will be relayed to the COC for analysis. Between ISS over-flights, internal diagnostics will be performed, typically once per day to make sure the systems are functional. Spare GLS units will be deployed as needed to maintain adequate coverage during the 3-5 year mission.

One GLS, dubbed the GLS-AXHL, that includes flashers and a horizontally pointing laser, will be installed in a P3B airplane managed by the NASA Airborne Science Program and operated out of NASA/WFF. The P3B features include an upward viewing portal that is available to install a GLS and a side port that will be fitted with a fused silica window to transmit the UV laser pulses. Once the necessary mechanical and electrical interfaces have been developed and tested, the airplane will be deployed monthly for under flights of the ISS at night. The P3B will fly out 500 km from the eastern seaboard to rendezvous with the ISS. Each flight will target a specific set of conditions. Over the length of the mission, these flights will cover a range of altitudes, atmospheric and cloud conditions, and moonlight. The flights will provide sufficient testing for the large number of events acquired by JEM-EUSO over the oceans. After each flight the GLS-AXHL will be dismounted and stored at WFF until the next flight. Details of the P3B flight data will be obtained and included in the analysis of the GLS-AXHL and JEM-EUSO data.

In addition to validating the accuracy of the EAS reconstruction analysis, the GLS tests the focus of the optical system. The GLS serves as a point source and the images of the GLS are small spots characteristic of the optics point-spread function. During the course of the mission, the GLS spot-size will be analyzed to determine if the telescope is maintaining proper focus. The downloaded images of the GLS will be analyzed and compared with optic simulations to guide any focusing adjustments required. Commands will be sent to the telescoping mechanism on JEM-EUSO to adjust the distance between the focal surface and optics to improve the focus.

## 4. Prototype EUSO Balloon Flight Tests

Balloon flights are planned to demonstrate a prototype of the JEM-EUSO telescope. The first flight is tentatively scheduled for 2014 from Timmins, Ontario. The key objectives are:

- A full-scale end-to-end test of JEM-EUSO technique
- Raising of the technology readiness level of key electronic components including HV control and onboard trigger
- Acceptance of signals over a large dynamic range
- Experimental determination of the effective UV background as seen looking down from 40 km altitude with the Balloon EUSO camera
- Acquisition  and processing of JEM-EUSO type data
- Detection of GLS-AXHL calibration signals from near-space
- Test the trigger algorithms with real and GLS-AXHL data
- The first imaging of an EAS looking down on the Earth's atmosphere

A prototype of the GLS-AXHL will be deployed to support the suborbital flights. In addition to testing the prototype and gaining experience with conducting under-flights, the GLS-AXHL will provide critical data for the above listed objectives. The balloon flights will be used to develop and verify components of the GLS design for the space mission and to exercise key elements of the GLS functions, operations and the analysis.

The GLS support for the balloon flights include a ground-based laboratory measurement of the balloon prototype's response and an airborne under-flight with the GLS-AXHL prototype. These exposures will be used to develop GLS design and to exercise key GLS functions, operations, and analysis procedures in preparation for supporting the space mission. Since the balloon will travel more slowly than the aircraft (unlike the ISS), the aircraft can fly multiple passes near the balloon to conduct these tests. The xenon flasher of the GLS prototype will be taken to the instrument integration site in France. After completing the integration of Balloon EUSO, ground tests will be conducted to assess its overall performance using the flasher. These tests will verify that the flash intensity and duration are compatible with the onboard trigger algorithm and the dynamic range of Balloon EUSO's focal surface detector. These test exposures will use the 4 optical filters planned for the space mission. Several alternate intensities will be tested.  The flasher will be tested over its ±30 degree field of view. The flasher will also be used to test Balloon EUSO over its ±6 degree field of view. Preliminary analysis of the acquired data will be completed immediately and a more in depth analysis will be completed after the instrument checkout period has been completed. This preliminary exposure will be used to understand the response of Balloon EUSO and improve the design of the xenon flasher for the balloon under flight to insure successful operations during the balloon mission.

The GLS-AXHL prototype will be taken to the balloon launch site well in advance of the flight so that it can be integrated into the chase aircraft which will support the balloon flight operations. A test flight with the GLS-AXHL prototype will be conducted to gain experience with under-flight operations before the balloon launch. Pre-flight and post-flight ground tests of the GLS and laser will be conducted to insure the sources are operational and meet the requirements for the balloon under-flight test.

The first JEM-EUSO prototype balloon flight will be launched from Timmins, Ontario, Canada in the spring or fall of 2014. After the balloon reaches its float altitude and the sun is setting on the balloon, the chase aircraft, with the GLS prototype onboard, will takeoff and establish communications with the Balloon Operations Center (BOC). The aircraft will use the current position of the balloon, as reported by the onboard GPS receiver, to fly to the balloon at night. If necessary, the aircraft can also use the radio beacon on the balloon to locate it. Once near the balloon and in communication with the BOC, the chase aircraft will execute an oval flight pattern at preselected altitudes. One leg of the oval will take the aircraft across the ±6° FoV of the JEM-EUSO prototype, which will be looking down to the nadir from the balloon and viewing a ~8 km diameter circle on the Earth below. During such passes, Xe flasher will be operated. On the other side of the oval, ~10 km distant from the balloon's ground track, the laser will shoot through the FoV. The BOC will provide feedback to the airborne personnel in near real-time on their attempts to detect the Xe flasher and laser signals. The aircraft will continue operations under the balloon until adequate calibration data are acquired. This balloon mission provides a unique opportunity to exercise and test several elements of the GLS system and data analysis in preparation for the space mission. In particular the balloon flight test enables the investigation of the atmospheric absorption in a single location over the range of altitudes spanned by the maxima of EASs.

## 5. Summary

The JEM-EUSO mission is planned for a three to five year mission following deployment on the ISS. JEM-EUSO must trigger on the UV light from EASs, measure the intrinsic luminosity of each EAS near its maximum and accurately reconstruct the EAS path so that the arrival direction of the initiating EECR can be found.

The GLS has been designed to provide 'ground truth' on the ability of the space-based JEM-EUSO camera to measure the intrinsic luminosity of EASs and determine the arrival direction of the initiating EECR. The GLS will be used to test and tune JEM-EUSO and its data processing algorithms to improve the accuracy of its measurements and to tune its trigger algorithm for detection of EECRs under a variety of atmospheric and background light conditions.

GLS units will be located in remote areas, free from anthropogenic background light, and at altitudes which place them above the planetary boundary layer, and within the range of altitudes at which the maxima of EASs occur. These remote sites will be chosen to sample the range of weather conditions around the World during which JEM-EUSO will measure EASs. In addition to the ground-based GLS units, one airborne unit is planned. This GLS unit will be used to conduct monthly under-fights of JEM-EUSO over the North Atlantic at a range of altitudes and weather conditions in order to conduct these calibrations over the open ocean.

All thirteen GLS systems will have a Xe flash lamp that will produce precisely calibrated and uniform flashes which are tailored to trigger JEM-EUSO. A laboratory prototype of this flasher has already been tested. In addition, seven of the GLS units, including the airborne one, will have a laser which will be used to simulate EASs. These calibration signals will be used to optimize the JEM-EUSO trigger and to test and improve the reconstruction of the arrival directions of the EECRs. The laser subsystem is based on the one currently in use at the Pierre Auger Observatory in Argentina.

A prototype GLS system, including both the Xe flasher and the laser, will be used during flights of the balloon-borne JEM-EUSO prototype, explore atmospheric absorption versus altitude and gain experience with aircraft operations. The first balloon flight is planned for 2014 from Timmins, Ontario, Canada.